# SOXS: Effects on optical performances due to gravity flexures, temperature variations, and subsystems alignment

Ricardo Zanmar Sanchez*[a], Matteo Aliverti[c], Matteo Munari[a], Sergio Campana[c], Riccardo Claudi[d], Pietro Schipani[e], Andrea Baruffolo[d], Sagi Ben-Ami[b,h], Federico Biondi[s], Giulio Capasso[e], Rosario Cosentino[g,a], Francesco D'Alessio[i], Paolo D'Avanzo[c], Ofir Hershko[h], Hanindyo Kuncarayakti[j,k], Marco Landoni[c], Giuliano Pignata[m], Adam Rubin[f], Salvatore Scuderi[a], Fabrizio Vitali[i], David Young[r], Jani Achrén[l], José Antonio Araiza-Duran[m], Iair Arcavi[n], Anna Brucalassi[m], Rachel Bruch[h], Enrico Cappellaro[d], Mirko Colapietro[e], Massimo Della Valle[e], Marco de Pascale[d], Rosario di Benedetto[a], Sergio D'Orsi[e], Avishay Gal-Yam[h], Matteo Genoni[c], Marcos Hernandez[g], Jari Kotilainen[j], Gianluca Li Causi[q], Seppo Mattila[k], Michael Rappaport[h], Kalyan Radhakrishnan[d], Davide Ricci[d], Marco Riva[c], Bernardo Salasnich[d], Stephen Smartt[r], Maximilian Stritzinger[o] and Hector Ventura[g]

[a]INAF - Osservatorio Astrofisico di Catania, Catania, Italy, [b]Harvard-Smithsonian Center for Astrophysics, Cambridge, USA, [c]INAF - Osservatorio Astronomico di Brera, Merate, Italy, [d]INAF - Osservatorio Astronomico di Padova, Padua, Italy, [e]INAF - Osservatorio Astronomico di Capodimonte, Naples, Italy, [f]European Southern Observatory, Garching, Germany, [g]INAF - Fundación Galileo Galilei, Breña Baja, Spain, [h]Weizmann Institute of Science, Rehovot, Israel, [i]INAF - Osservatorio Astronomico di Roma, Rome, Italy, [j]FINCA - Finnish Centre for Astronomy with ESO, Turku, Finland, [k]University of Turku, Turku, Finland, [l]Incident Angle Oy, Turku, Finland, [m]Universidad Andres Bello, Santiago, Chile, [n]Tel Aviv University, Tel Aviv, Israel, [o]Aarhus University, Aarhus, Denmark, [q]INAF - Istituto di Astrofisica e Planetologia Spaziali, Rome, Italy, [r]Queen's University Belfast, Belfast, UK, [s]Max-Planck-Institut für Extraterrestrische Physik, Garching, Germany

## ABSTRACT

SOXS (Son Of X-Shooter) is the new medium resolution wide-band spectrograph to be installed at the 3.6m New Technology Telescope (NTT) in La Silla. SOXS will offer simultaneous wavelength coverage from 0.35 to 2.0 μm and will be dedicated to the study of transient and variable sources. While nominal optical performances of the system were presented in previous proceedings (Zanmar Sanchez et al. 2018), we here present a set of further analyses aimed to identify and quantify optical effects, due to changes in temperature and orientation of the instrument during alignment and operations.

## 1. INTRODUCTION

SOXS (Son of X-Shooter) is a medium resolution (R~5000) spectrograph, currently under manufacturing and integration, to be installed on the NTT Telescope at la Silla[1]. It is expected to start operations in 2022 and it was designed to observe the wavelength range from 350 to 2000 nm using two different spectrographs: the UVVIS (350-850 nm) and NIR (800-2000 nm). The overall optical design of SOXS was presented in a previous paper[2]. Briefly, SOXS consists of a Common Path (CP) that splits and relays incoming light from the Telescope into the NIR and UVVIS spectrographs. The wavelength separation is done with a dichroic and both arms of the CP have a tip/tilt mirror and lenses to change the F/#. The relative focus between the two CP arms can be adjusted by moving the doublet inside the NIR arm and the tip/tilt mirrors are used to make adjustments such as the ones presented in this work.

*ricardo.sanchez@inaf.it; phone +39 095 733-2206

In this work we present further analysis that ensure that the design is robust and maintains the image quality and performance under the expected temperature variations and gravity flexures. We also explore the range of tolerance needed during the alignment of a couple of subsystems.

This document is organized as follows: we start by describing the gravitational effects on the spectrograph for 8 different directions and present the effect on image quality and movement of the image on the focal plane; we then describe the effects on image quality due to temperature variations. Finally, we explore the tolerances on decenter and tilts during alignment of Telescope, Common Path and NIR spectrograph.

## 2. EFFECT OF GRAVITY FLEXURES ON OPTICAL PERFORMANCE

An Ansys[3] analysis was performed in order to estimate the deformations suffered due to gravity by the flange and all the mountings that support the SOXS spectrograph optics. Since SOXS is expected to rotate in the altazimuth mount of the NTT Telescope, a total of 8 different cases have been simulated with the gravity pointing in different directions as illustrated in Figure 1. Below we consider the effects of flexures on both the image position on the focal plane of the CP and on the NIR spectrograph detector.

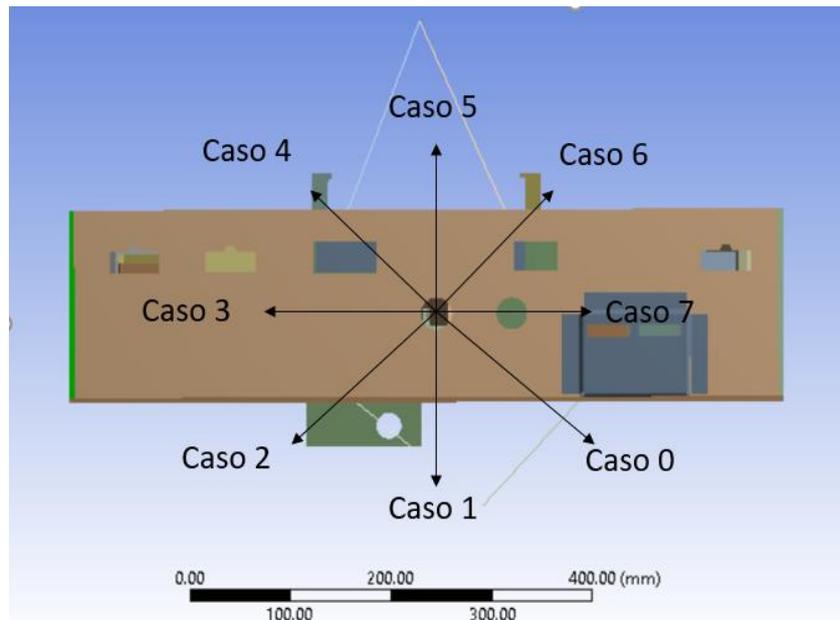

Figure 1. A side view of the Common Path enclosure with arrows representing the direction of gravity for the eight simulated cases.

### 2.1 Image position on the slit

For each gravity direction and for each optical element in the Common Path, the resulting deformations were expressed as tilts and displacements of elements in our optical design. In Figure 2 we present an example of the deformations for the NIR doublet mounting[4] inside the CP-NIR and its reference frame. After applying these tilts and decenters to the nominal optical design, the image spot centroid in the slit moves in X (short side of the slit), Y (long side of the slit) and Z (focus) with respect to the nominal position. These values are reported for all 8 gravity cases in Table 1 and Figure 3 for both the NIR and UVVIS arms of the CP. From Figure 3, the maximum image offsets that we need to correct for are about 10 microns. Since the CP response to flexures is linear, even assuming gravity flexures twice as large (to account for possible errors in the analysis), the resulting maximum offsets of 20 microns are well within the designed capabilities of the tip/tilt mirrors of both arms of the CP. Note that along Z there are no effects due to flexures (the system remains in focus).

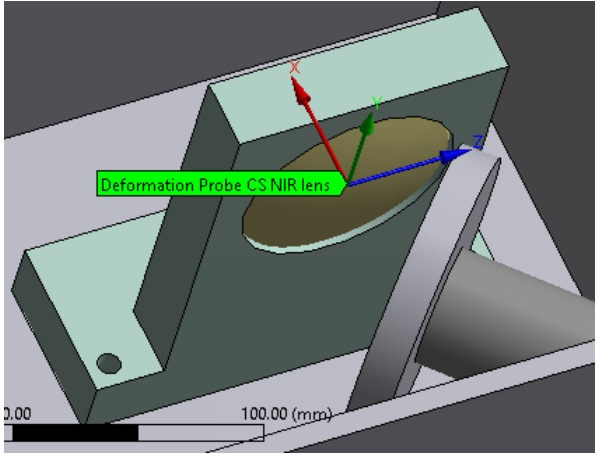 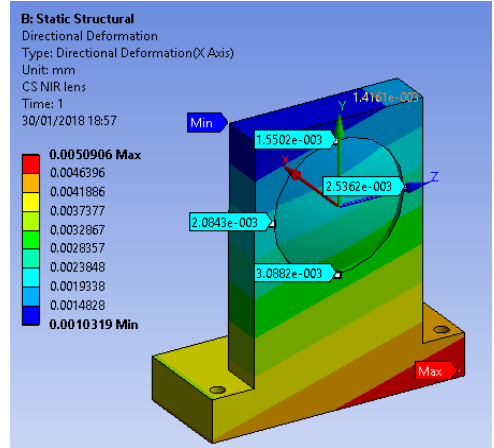

Figure 2. *Left*, the mounting for the NIR doublet in the Common Path, *right*, the directional deformation of it due to gravity.

Table 1. NIR and UVVIS image offset from nominal position on the slit.

|  | Case 0 | Case 1 | Case 2 | Case 3 | Case 4 | Case 5 | Case 6 | Case 7 |
|---|---|---|---|---|---|---|---|---|
|  | **CP-NIR** | | | | | | | |
| dx (mm) | 2.0E-03 | 6.9E-03 | 2.7E-03 | -2.0E-03 | -6.9E-03 | -6.9E-03 | -2.7E-03 | 2.0E-03 |
| dy (mm) | 7.6E-03 | -6.7E-03 | -1.0E-02 | -7.6E-03 | -6.4E-04 | 6.7E-03 | 1.0E-02 | 7.6E-03 |
| dz (mm) | 0 | 0 | 0 | 0 | 0 | 0 | 0 | 0 |
|  | **CP-UVVIS** | | | | | | | |
| dx (mm) | 2.0E-03 | 1.0E-02 | 5.7E-03 | -2.0E-03 | -8.5E-03 | -1.0E-02 | -5.7E-03 | 2.0E-03 |
| dy (mm) | -3.6E-06 | 6.7E-03 | 3.8E-03 | -1.4E-03 | -5.7E-03 | -6.7E-03 | -3.8E-03 | 1.4E-03 |
| dz (mm) | 0 | 0 | 0 | 0 | 0 | 0 | 0 | 0 |

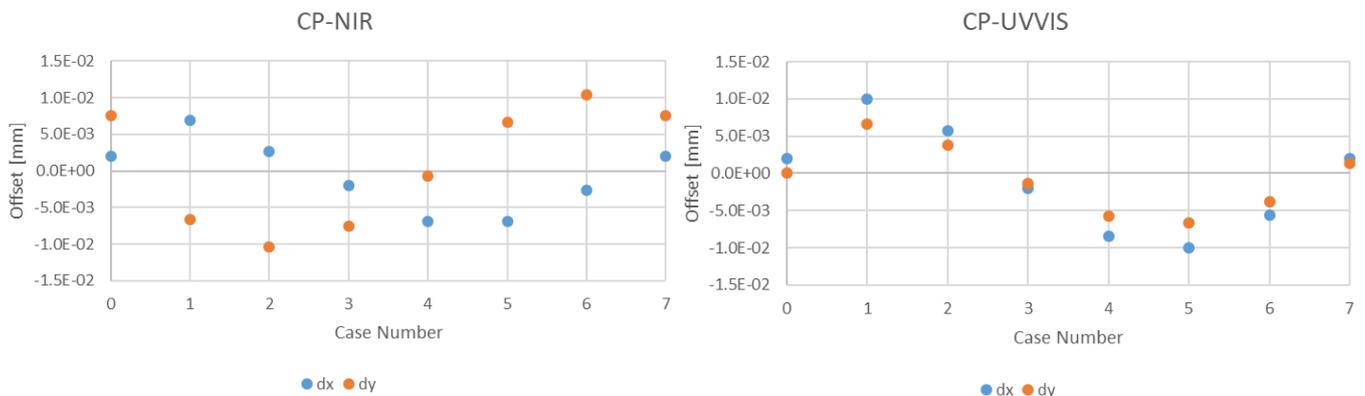

Figure 3. CP-NIR (*left*) and CP-UVVIS (*right*) image offsets on the slit from nominal position. X axis is the simulated gravity case number and the Y axis is the offset in mm along the short (blue) and long (orange) side of the slit.

## 2.2 Image position on the NIR detector

As for the Common Path, we now consider the effects of gravity flexures on the spectra imaged on the NIR detector. To that end, we take the deformations found by the Ansys analysis and express them in terms of tilts/displacements of optical elements inside the NIR spectrograph in our optical design. Note that the slit was assumed to be correctly illuminated by the CP, e.g. the tip/tilt mirror has already corrected gravity flexures inside the NIR arm of the Common Path. The resulting movement of the spectrum on the detector along the dispersion (blue line) and spatial (orange line) directions are summarized in Figure 4 for different rotation angles of the spectrograph.

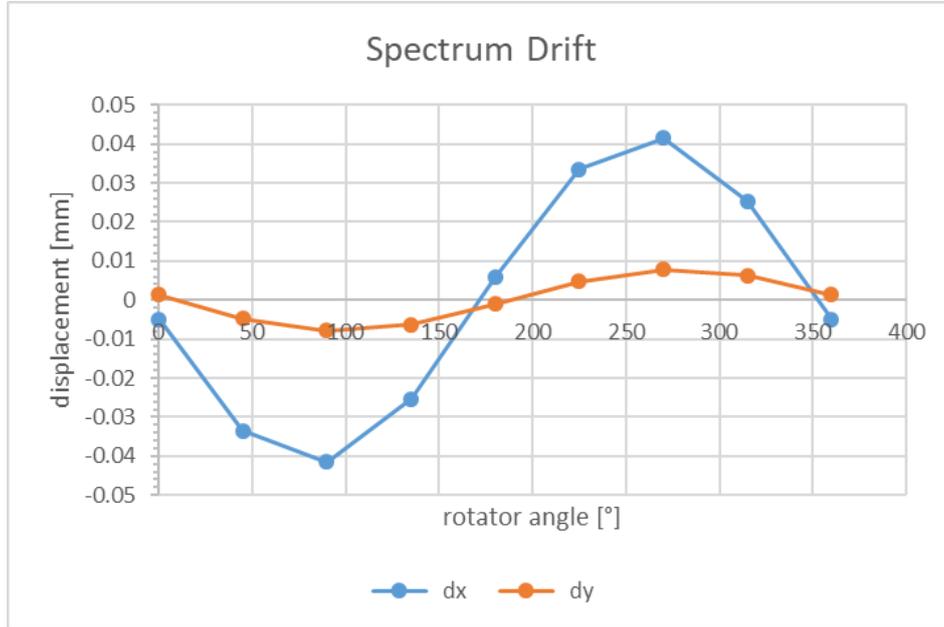

Figure 4. Spectrum drift in mm with respect to nominal position. In blue, the drift of the spectrum in the dispersion direction; in orange, the drift in the spatial direction.

In the plot, we see that if the spectrograph were to rotate by 180º during an observation, drifts of up to 4.5 pixels (18 microns = 1 pixel) are possible. Fortunately, when we consider the expected exposure times needed by the NIR spectrograph and the pointing direction of the Telescope, we will rarely be in such a large drift situation as we can see in the following analysis.

For an altazimuth Telescope, such as the NTT, located at latitude $\varphi$, the field rotates by Q degrees during an exposure time T according to the following expression[5]: $Q = \omega_0 T \cos(\varphi) \cos(A) / \cos(a)$, where A and $a$ are the target azimuth and altitude respectively, and $\omega_0$ is Earth's sidereal rate. Note that the expression implies large field rotations (e.g. large drifts on the detector) only when observing near the zenith. For example, in Table 2 we present the expected drift in pixels, along the dispersion direction of the spectrum on the NIR detector, for a 30 min exposure for different pointed azimuth and altitude angles.

Moreover, we can calculate what percentage of sky will be affected by high rotation angles (taking into account the spherical area of the Alt/Az bin). Assuming that the Telescope will be pointed uniformly in the range Alt,Az=[20°-87°,0-360°] which is accessible to the NTT Telescope, Table 3 gives the percentage of the sky affected for a few different exposure times and minimum pixel drifts.

Table 2. Pixel drift on the detector for a 30 min exposure for different azimuth/altitude angles.

| Alt[°] | 21 | 31 | 41 | 51 | 61 | 71 | 81 | 87 |
|---|---|---|---|---|---|---|---|---|
| Az [°] | | | | | | | | |
| 5 | 0.3 | 0.3 | 0.4 | 0.4 | 0.6 | 0.9 | 1.8 | 4.4 |
| 15 | 0.3 | 0.3 | 0.4 | 0.4 | 0.6 | 0.8 | 1.7 | 4.3 |
| 25 | 0.3 | 0.3 | 0.3 | 0.4 | 0.5 | 0.8 | 1.6 | 4.1 |
| 35 | 0.2 | 0.3 | 0.3 | 0.4 | 0.5 | 0.7 | 1.5 | 3.8 |
| 45 | 0.2 | 0.2 | 0.3 | 0.3 | 0.4 | 0.6 | 1.3 | 3.4 |
| 55 | 0.2 | 0.2 | 0.2 | 0.3 | 0.3 | 0.5 | 1.0 | 2.9 |
| 65 | 0.1 | 0.1 | 0.2 | 0.2 | 0.2 | 0.4 | 0.8 | 2.2 |
| 75 | 0.1 | 0.1 | 0.1 | 0.1 | 0.2 | 0.2 | 0.5 | 1.4 |
| 85 | 0.0 | 0.0 | 0.0 | 0.0 | 0.1 | 0.1 | 0.2 | 0.5 |
| 95 | 0.0 | 0.0 | 0.0 | 0.0 | -0.1 | -0.1 | -0.2 | -0.5 |
| 105 | -0.1 | -0.1 | -0.1 | -0.1 | -0.2 | -0.2 | -0.5 | -1.4 |
| 115 | -0.1 | -0.1 | -0.2 | -0.2 | -0.2 | -0.4 | -0.8 | -2.2 |
| 125 | -0.2 | -0.2 | -0.2 | -0.3 | -0.3 | -0.5 | -1.0 | -2.9 |
| 135 | -0.2 | -0.2 | -0.3 | -0.3 | -0.4 | -0.6 | -1.3 | -3.4 |
| 145 | -0.2 | -0.3 | -0.3 | -0.4 | -0.5 | -0.7 | -1.5 | -3.8 |
| 155 | -0.3 | -0.3 | -0.3 | -0.4 | -0.5 | -0.8 | -1.6 | -4.1 |
| 165 | -0.3 | -0.3 | -0.4 | -0.4 | -0.6 | -0.8 | -1.7 | -4.3 |
| 175 | -0.3 | -0.3 | -0.4 | -0.4 | -0.6 | -0.9 | -1.8 | -4.4 |

For example, for a 600 s observation, the spectrum would drift by 1 pixel or more, only in 0.3% of the observable area of the sky by NTT. Given that the SOXS NIR spectrograph is not expected to observe continuously for more than 10 minutes due to its nodding strategy, the percentage of the sky where we will suffer a drift larger than half a pixel is very small.

Table 3. Percentage of the observable sky affected by spectrum drift on the detector for different exposure times.

| Exposure Time [s] | minimum pixel drift | | | | |
|---|---|---|---|---|---|
| | >0.5 | >0.75 | >1 | >1.5 | >2 |
| **600** | 1.2% | 0.6% | 0.3% | 0.1% | 0% |
| **900** | 3% | 1.2% | 0.7% | 0.3% | 0.1% |
| **1800** | 12.6% | 5.6% | 3% | 1.2% | 0.7% |
| **3600** | 57% | 24% | 13% | 5% | 3% |

## 3. EFFECTS OF TEMPERATURE VARIATIONS ON IMAGE POSITION

In contrast with the NIR spectrograph which is temperature controlled, the Common Path is expected to experience temperature variations from 0-20°C throughout the year. Therefore, here we study the thermal behavior of this subsystem to see the effects on the image position on the NIR slit. The final goal is to estimate the differences in position of best focus and slit positions, to understand if temperature changes may cause defocus and/or wandering on the slit plane of the image at the entrance of the spectrographs, and to discuss the possible rotation of the spectrographs with respect to CP slit orientation.

Figure 5 presents the three subsystems considered for thermal expansion: Telescope flange (amber), the Common Path (green) and the NIR spectrograph (red). The adopted coordinate system is Y upwards along the incoming light from the Telescope, X to the right towards the UVVIS spectrograph (not shown) and Z out of the page. The spectrographs and Common Path are fixed to the flange with kinematics mounts (KM). In fact, by careful placement of the KM between the spectrograph and the flange[4], we can make the NIR slit co-move with the flange during the thermal expansion/contraction as explained below.

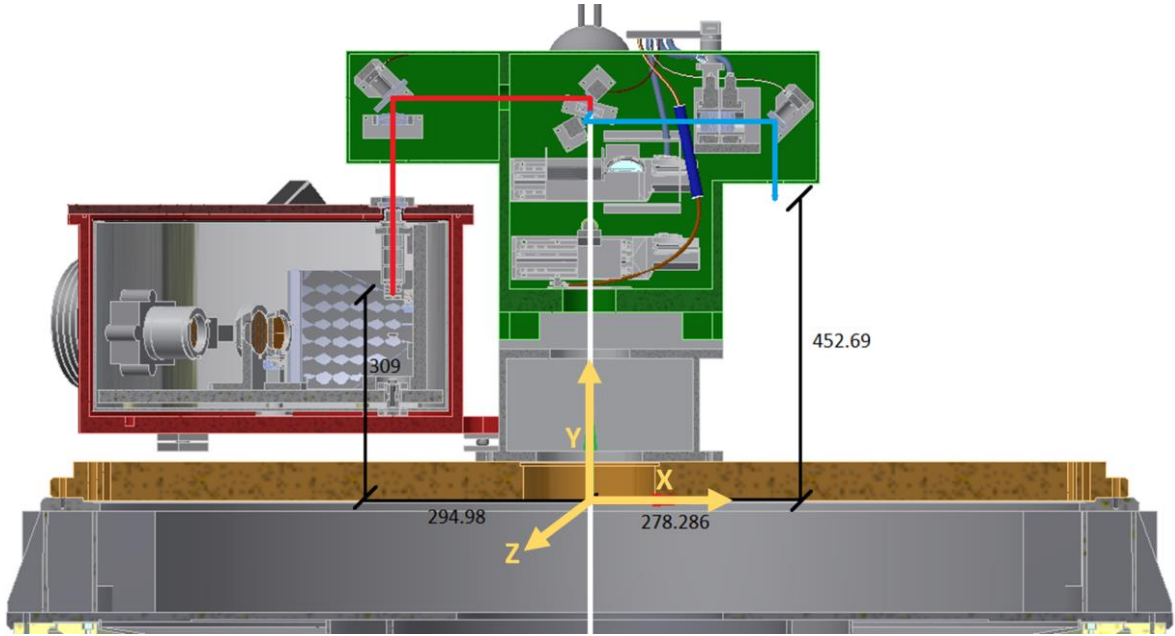

Figure 5. Sketch of the instrument with dimensions. The UVVIS spectrograph is not shown. The flange (amber) and the Common Path (green) are independently expanded due to thermal variations and we consider the consequences in terms of image position on the slit and focus.

Since we require the NIR slit, which is inside the spectrograph, to co-move with the flange as the temperature changes, we place the spherical KM, that constraints 3 degrees of freedom, below the slit. We also require the long side of the slit to be parallel to the radial direction of the flange to avoid rotation, therefore the cylindrical KM, which constrains 2 degrees of freedom, is aligned with the radial direction of the flange. The third KM, a spherical washer that constraints the last degree of freedom, can be placed anywhere.

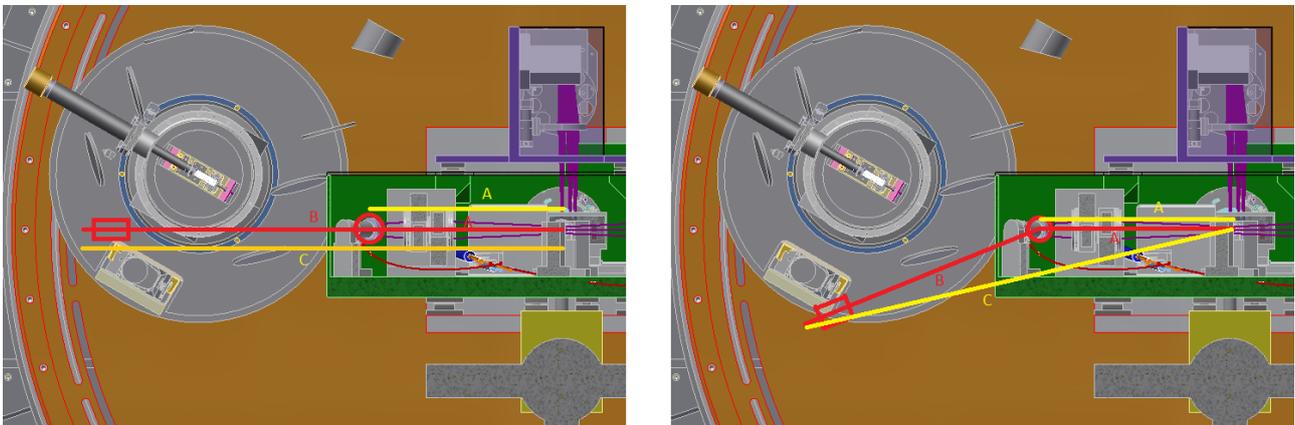

Figure 6. Positions of kinematic mounts. The circle represents the spherical KM, the rectangle is the cylindrical KM. *Left*, the correct adopted placement of the KM where the spherical KM coincides with the position of the slit and the cylindrical KM is parallel to the radial direction of the flange. *Right*, a hypothetical incorrect placement of the KM that would induce rotation of the slit during temperature changes.

On the left of Figure 6 we illustrate the adopted KM concept for the NIR spectrograph attachment to the flange and on the right, a hypothetical incorrect concept that would induce slit rotation as the flange expands radially. With the adopted solution, we only need to consider the relative movement between the flange and the Common Path since the slit and flange will co-move. Below we independently consider the thermal expansion of the flange and then the thermal expansion of the Common Path and check the resulting relative movement of the image on the slit.

### 3.1 Flange Deformations

To estimate the deformation of the flange, an Ansys analysis has been carried out. The analysis reflects the foreseen configuration: an aluminum flange bolted at the border to a steel structure. Both steel and aluminum may expand/contract with temperature variations. Figure 7 depicts the aluminum flange stress variations (left) and the corresponding total deformations (right) due to temperature variations.

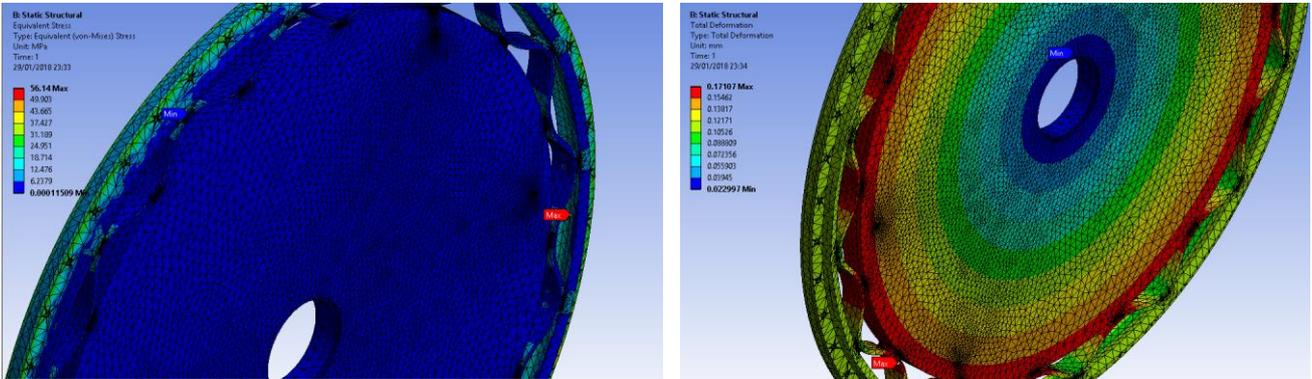

Figure 7. *Left*, stresses on the flange. *Right*, total deformation of the flange (in all directions) due to temperature variations.

The adopted parameters for the model are summarized in Table 4. The analysis is essentially a linear differential analysis, and although it was performed for ΔT = -10 °C, results may be rescaled linearly for any similar range of temperatures.

Table 4. Adopted model parameters in Ansys for the flange deformation modeling.

| CTE Aluminum | 23E-6 K-1 |
|---|---|
| CTE Steel | 14E-6 K-1 |
| ΔT | -10 °C |
| Flange Radius | 795 mm |

In Figure 8 we present the directional deformation along Y (focus direction) and along the radial direction (e.g. parallel to the slit). Note that flange flexures colors are greatly magnified for visualization purposes. We find that in the focus direction, the flange expands or contracts by 12±5 microns for a ΔT=10ºC. Along the radial direction we find that both NIR and UVVIS spectrograph slits would move by ~70 microns for a ΔT=10ºC. Our analysis also finds that there are no torsional deformations of the flange.

We summarize the resulting change in position due to thermal expansion of mechanical elements for ΔT=+10°C in Table 5.

Table 5. Change in position of the projections on the flange of the center of the UVVIS and NIR slits.

|  | UVVIS | NIR |
|---|---|---|
| ΔX [microns] | 64 | -70 |
| ΔY [microns] | 12 | 12 |
| ΔZ [microns] | 0 | 0 |

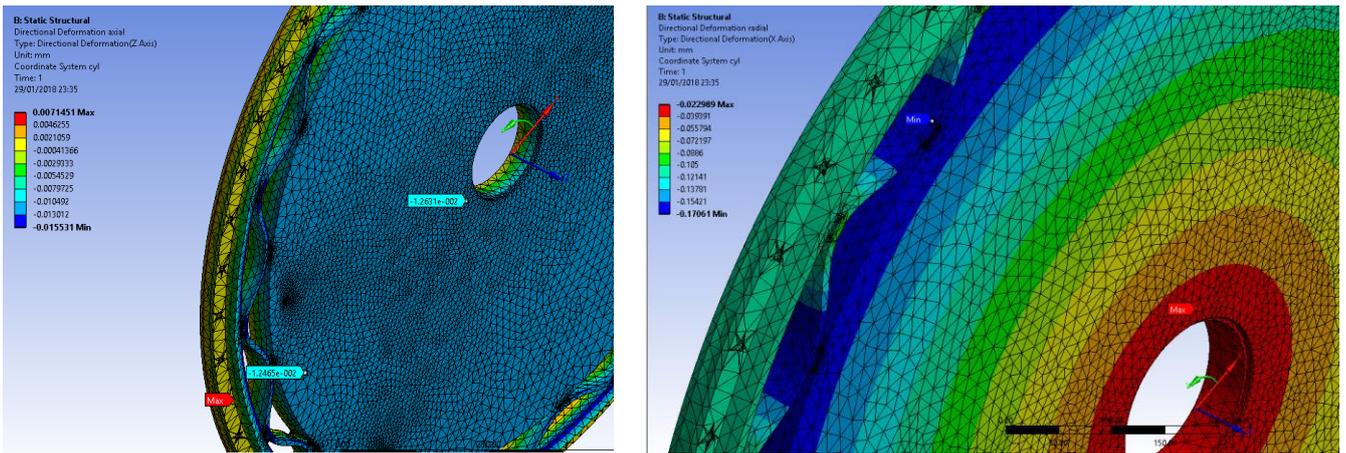

Figure 8. *Left*, deformations along focus axis (Y): image shows a uniform shift of the flange with temperature, with a variation of +0.012 mm for ΔT=+10°C. *Right*, deformations on the flange plane along the radial direction. The estimated variation for the position corresponding to both slits (green area) is about 0.070mm.

### 3.2 Optical simulations

We now consider the thermal expansion of the Common Path in a ray-tracing software. For the optical simulations we adopted the nominal position of the back of the flange center as Global Coordinate reference. The system was evaluated at 5 temperatures: 0, 5, 10, 15, and 20°.

The Y displacement of the flange, due to temperature variations was accounted for by inserting a surface of variable thickness according to the results of the previous section. This has the effect of displacing the entire CP with respect to the nominal position of the flange. By design, the Common Path structure is made out of the same aluminum as the flange, therefore we adopted a CTE= $23\times10^{-6}$ K$^{-1}$ for all the distances between optical elements up to last mirror in UVVIS and the last lens in the NIR arm (Figure 9). The resulting thermal expansion is summarized in Table 6. Note that the relative defocus between the NIR and UVVIS arms was adjusted by moving the doublet in the NIR arm of the CP.

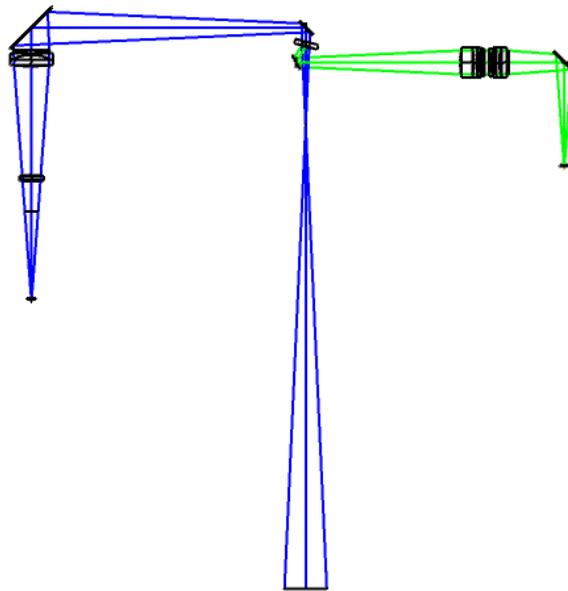

Figure 9. Common Path optical layout from flange interface to NIR (blue) and UVVIS (green) focal planes.

Table 6. Thermal expansion by UVVIS and NIR arms of the CP. Y is in the direction of focus. NIR doublet adjustment to remain in focus is reported. All units are in mm.

|   |   | 0° | 5° | 10° | 15° | 20° |
|---|---|---|---|---|---|---|
| **UVVIS** | ΔX | -0.059 | -0.030 | 0.000 | 0.030 | 0.059 |
|   | ΔY | -0.015 | -0.007 | 0.000 | 0.007 | 0.013 |
|   | ΔZ | 0.000 | 0.000 | 0.000 | 0.000 | 0.000 |
|   |   |   |   |   |   |   |
|   |   | 0° | 5° | 10° | 15° | 20° |
| **NIR** | ΔX | 0.068 | 0.034 | 0.000 | -0.034 | -0.068 |
|   | ΔY | -0.015 | -0.007 | 0.000 | 0.007 | 0.014 |
|   | ΔZ | 0.000 | 0.000 | 0.000 | 0.000 | 0.000 |
| NIR Doublet | ΔY | 0.10 | 0.05 | 0.00 | -0.05 | -0.1 |

From these results, we find that the position of the best focus remains stable in Y (focus) position in a range of 15 microns for the UVVIS, and similar figures for the NIR (once the NIR lens is moved by +/- 100 micron). We stress that these figures are obtained assuming that the Telescope Focal Plane best position remains fixed in the global coordinates system, at (0,500,0) mm from the nominal flange center. We have checked the optical quality of the image with spot diagrams and they remain essentially constant for different temperatures.

### 3.3 Results: image displacement on the slit.

Given the results presented above in Table 5 and Table 6 we conclude that on the plane XZ (parallel to the flange) the image and the flange itself will expand by about the same values: 0 mm in tangential directions and ~0.07 mm in the radial direction. Since by design the slit is co-moving with the flange, no movement of the image on the slit will be present. If we consider a 10-20% error in analyses due to CTE inhomogeneity or differences in flange and CP, modelling, etc., we can expect a possible ~15 microns movement along the long side of the slit (1350 microns) which can be adjusted with the tip/tilt mirror of the CP. On the Y direction, focus remains in the nominal position in +/-15 micron range for both arms (after refocusing NIR CP). This takes into account a +/-20micron displacement of the flange which is compensated by the expansion of the CP.

## 4. EFFECTS OF SUBSYSTEMS ALIGNMENT ON OPTICAL QUALITY

During the assembly process of SOXS, different subsystems will have to be integrated and aligned[6]. In this section we aim to find tolerances for the maximum allowed decenter and tilt between subsystems that would still preserve acceptable image quality. Below we model the alignment between Telescope and Common Path and between Common Path and NIR spectrograph.

### 4.1 Alignment between Telescope and CP

The interface between the Telescope and CP is at the flange, therefore we introduce decenters and tilts there as indicated in Figure 10 where the reference frame is also indicated. Starting with the nominal subsystems, where alignment is perfect, we introduce separately tilts and decenters at the flange interface at regular intervals for a large range of values. For each tilt or decenter, we evaluate the image quality as the radius of the fractional encircled energy of 80% (Zemax[7] GENC operator). Some of the probed tilts and decenters will be large enough to produce vignetting on the focal plane of the CP.

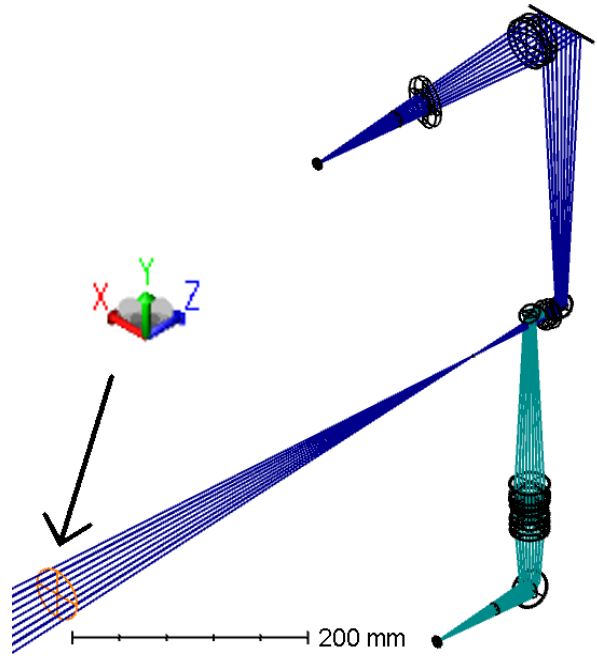

Figure 10. Reference frame for CP. Arrow indicates the position of the flange, interface between Telescope and CP.

An example of the results for the UVVIS decenter is presented in Figure 11 where we plot the percentage difference of GENC with respect to the nominal value produced by decenters in X (left) and in Y (right). The whole range in the X axis represents the interval that does not produce vignetting in the focal plane of the CP. With horizontal lines we have marked when the percentage difference is 3%, 5% and 10% and these will provide the required tolerance intervals to keep during alignment. A summary of the results for both arms of the CP is presented in Table 7. NA values in the table simply mean that we start vignetting before being able to evaluate GENC. In general, we see that the requirements for the alignment are somewhat relaxed except for the decenter in Z (focus). Along this axis however there is a motorized sledge with enough precision[8].

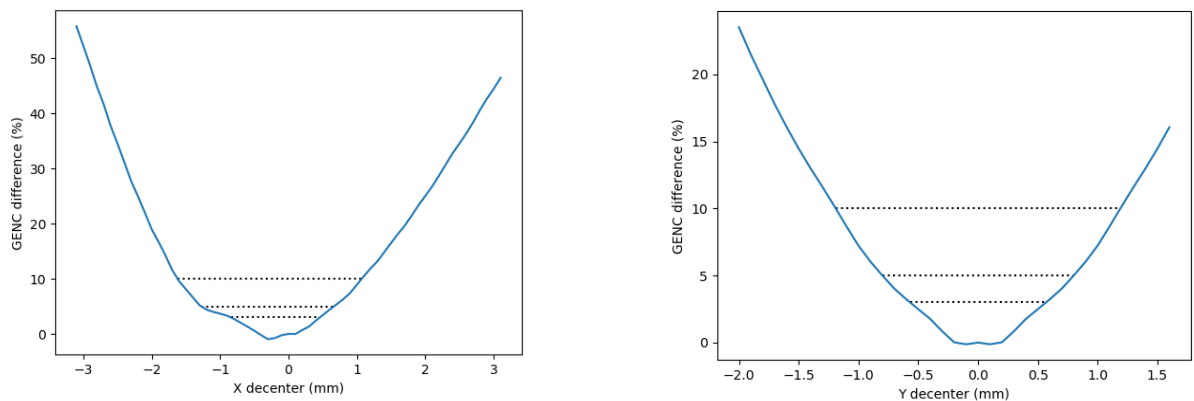

Figure 11. GENC percentage difference from nominal produced by decenters. The whole range in X axis is the non-vignetted interval.

Table 7. GENC percentage difference from nominal produced by tilts/decenters between CP-NIR/CP-UVVIS and Telescope.

| | NIR | | | | | | | |
|---|---|---|---|---|---|---|---|---|
| | 3% worst | | 5% worst | | 10% worst | | Non vignetted | |
| | min | max | min | max | min | max | min | max |
| **Tilt X (deg)** | na | 0.076 | na | na | na | na | -0.100 | 0.080 |
| **Tilt X (deg)** | -0.080 | na | -0.130 | na | na | na | -0.180 | 0.180 |
| **Dec X (mm)** | -0.689 | na | -1.135 | na | -2.370 | na | -3.100 | 3.100 |
| **Dec Y (mm)** | -0.657 | na | na | na | na | na | -0.900 | 1.500 |
| **Dec Z (mm)** | -0.068 | .090 | -0.088 | 0.142 | -0.138 | 0.190 | <-0.500 | > 0.480 |
| | UVVIS | | | | | | | |
| | 3% worst | | 5% worst | | 10% worst | | Non vignetted | |
| | min | max | min | max | min | max | min | max |
| **Tilt X (deg)** | -0.060 | 0.060 | -0.082 | 0.082 | -0.119 | 0.119 | -0.140 | 0.140 |
| **Tilt Y (deg)** | -0.068 | 0.057 | -0.094 | 0.075 | -0.144 | 0.114 | -0.220 | 0.220 |
| **Dec X (mm)** | -0.845 | 0.460 | -1.276 | 0.665 | -1.622 | 1.070 | -3.100 | 3.100 |
| **Dec Y (mm)** | -0.569 | 0.569 | -0.800 | 0.800 | -1.190 | 1.190 | -2.000 | 1.600 |
| **Dec Z (mm)** | -0.058 | 0.023 | -0.067 | 0.032 | -0.087 | 0.060 | <-0.500 | >0.480 |

Misalignment will also decenter and change the incidence angle (RAID Zemax operator) of the image in the focal plane of the CP. Given that the behavior is linear, we can summarize the results with fitted slopes. For example, we find that for both arms of the CP, a decenter in X or Y of 1mm between Telescope and CP, results in a decenter of 0.6 mm in the focal plane and an incidence angle of 0.013 deg (NIR) or ~ 0.03 deg (UVVIS). A tilt in X or Y of 1 deg results in a decenter of 5.2 mm and a RAID of 1.6 deg (NIR) or 1.3 (UVVIS). The results are summarized in Table 8.

Table 8. Resulting decenter/RAID produced by tilts/decenters for both CP-NIR and CP-UVVIS. The number is the fitted slope.

| | Effect on focal plane of CP-NIR | | | Effect on focal plane of CP-UVVIS | | |
|---|---|---|---|---|---|---|
| **Misalignment** | x | y | RAID | x | y | RAID |
| **X** | 0.6 [mm/mm] | | 0.013 [deg/mm] | 0.6 [mm/mm] | | 0.04 [deg/mm] |
| **Y** | | 0.6 [mm/mm] | 0.013 [deg/mm] | | 0.6 [mm/mm] | 0.03 [deg/mm] |
| **Tx** | | -5.1 [mm/deg] | 1.6 [deg/deg] | | -5.1 [mm/deg] | 1.3 [deg/deg] |
| **Ty** | 5.1 [mm/deg] | | 1.6 [deg/deg] | 5.1 [mm/deg] | | 1.3 [deg/deg] |

### 4.2 Alignment between CP and NIR spectrograph

We now simulate possible misalignments between the Common Path and NIR spectrograph and check the effects on the image quality of the observed spectrum. As before, misalignments are modelled in our optical design with decenters in X, Y and Z and tilts in X and Y at the interface between the CP and NIR spectrograph, adopted to be the slit. The reference frame is depicted in Figure 12 where the slit position is indicated with an arrow.

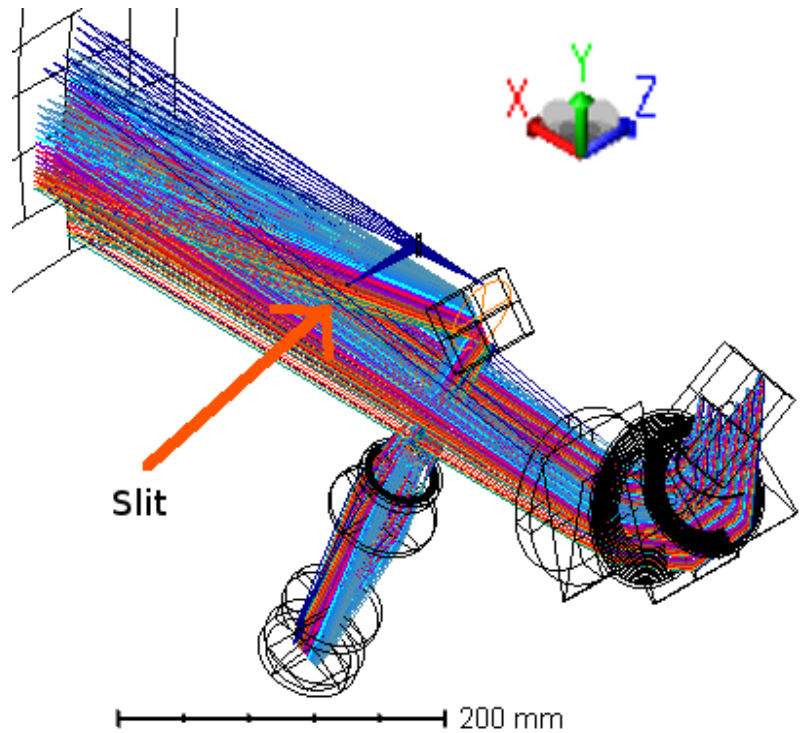

Figure 12. NIR spectrograph. Tilts and decenter were simulated around the slit indicated with an arrow.

As in the previous section, to check the image quality of the spectrum, we have chosen the radius of the fractional encircled energy at 80% (GENC) for 40 different wavelengths sampling the NIR wavelength range from 787 nm to 2010 nm. After we apply a decenter or tilt, we find the percentage difference between the perturbed and nominal GENC. An example of the results for tilts is in Figure 13, where we present the maximum percentage difference for *any* of the considered wavelengths as a function of the applied tilt X (left) and tilt Y (right). In the figure, we have marked with vertical lines the range of tilts that produce a 3%, 5% or 10% difference with nominal. The results for decenters and tilts are summarized in Table 9. In the table we also indicate the intervals for which there is no vignetting but note that not all the misalignments introduce vignetting, e.g. defocus (Z decenter). The requirement for decenter in Z (defocus) is particularly stringent but along this direction a sledge motor will be used with enough precision.

Table 9. GENC percentage difference from nominal produced by tilts/decenters.

|  | 3% worst | | 5% worst | | 10% worst | | Non vignetted | |
| --- | --- | --- | --- | --- | --- | --- | --- | --- |
|  | min | max | min | max | min | max | min | max |
| **Tilt X (deg)** | -0.200 | 0.075 | -0.316 | 0.147 | -0.573 | 0.213 | -0.4 | na |
| **Tilt Y (deg)** | -0.250 | 0.154 | -0.467 | 0.241 | -0.821 | 0.447 | -0.3 | 0.3 |
| **Dec X (mm)** | -0.383 | 0.213 | -0.635 | 0.351 | -1.204 | 0.845 | -0.8 | 1.0 |
| **Dec Y (mm)** | -0.075 | 0.067 | -0.093 | 0.110 | -0.210 | 0.205 | na | na |
| **Dec Z (mm)** | -0.006 | 0.005 | -0.010 | 0.008 | -0.020 | 0.015 | na | na |

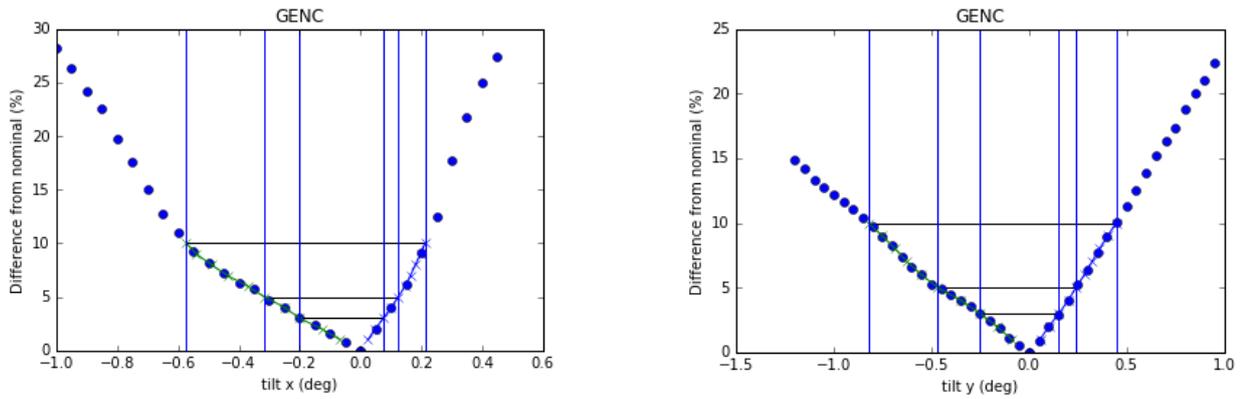

Figure 13. Maximum (of 40 wavelengths) percentage difference of GENC between nominal and tilt x (left) and tilt y (right) case. Similar plots where created for decenter X, Y and Z. Horizontal and vertical lines mark the 3%, 5% and 10% percentages used for tolerance.

## 5. CONCLUSIONS

With SOXS expected to start operations in the near future, new analysis was necessary to ensure that it will conform to the requirements. In this work we have presented thermal, gravity flexure analysis and tolerances on alignment of subsystems. We found that both the image on the slit and on the NIR spectrograph detector won't deviate much from the nominal case as SOXS rotates on the telescope. We also find that expected temperature variations on the CP won't have an impact on the image of the slit as small offsets can be corrected with the tip/tilt mirrors.